\documentclass[apjl]{emulateapj}  
\usepackage{graphicx}
\usepackage{multirow}
\usepackage{xcolor}

\usepackage{amsmath}
\usepackage{natbib}
\usepackage{hyperref}

\usepackage[normalem]{ulem}

\begin{document}

\title{On the maximum black hole mass at solar metallicity}

\author{
 Amedeo Romagnolo\altaffilmark{1}
 Alex C. Gormaz-Matamala\altaffilmark{1}
 Krzysztof Belczynski\altaffilmark{1}\altaffilmark{$\dagger$}
}

\affiliation{
   $^{1}$ Nicolaus Copernicus Astronomical Center, Polish Academy of Sciences, ul. Bartycka 18, 00-716 Warsaw, Poland
}

\altaffiliation{
    $^{\dagger}$ Deceased on 13th January 2024.
}

\email{amedeoromagnolo@gmail.com}

\begin{abstract}

In high metallicity environments the mass that black holes (BHs) can reach just after core-collapse widely depends on how much mass their progenitor stars lose via winds. On one hand new theoretical and observational insights suggest that early-stage winds should be weaker than what many canonical models prescribe. On the other hand the proximity to the Eddington limit should affect the formation of optically thick envelopes already during the earliest stages of stars with initial masses $M_{\rm ZAMS}\gtrsim 100$~M$_\odot$, hence resulting in higher mass-loss rates during the main sequence. We use the evolutionary codes {\tt MESA} and {\sc Genec} to calculate a suite of tracks for massive stars at solar metallicity Z$_\odot=0.014$ which incorporate these changes in our wind mass loss prescription. In our calculations we employ moderate rotation, high overshooting and magnetic angular momentum transport. We find a maximum BH mass $M_{\rm BH, max}=28.3$~M$_\odot$ at Z$_\odot$. The most massive BHs are predicted to form from stars with $M_{\rm ZAMS}\gtrsim 250$~M$_\odot$, with the BH mass directly proportional to its progenitor's $M_{\rm ZAMS}$. We also find in our models that at Z$_\odot$  almost any BH progenitor naturally evolves into a Wolf-Rayet star due to the combined effect of internal mixing and wind mass loss. These results are considerably different from most recent studies regarding the final mass of stars before their collapse into BHs. While we acknowledge the inherent uncertainties in stellar evolution modelling, our study underscores the importance of employing the most up-to-date physics in BH mass predictions.
\end{abstract}


\section{Introduction}
\label{sec.intro}

The LIGO/Virgo/KAGRA (LVK) collaboration has delivered summary of detections of compact object mergers in gravitational-waves in $3$ observing runs O1/O2/O3~\citep{LigoO3b}. The majority of these detections are black hole black hole (BH-BH) mergers. BHs are found in wide mass range $\sim 2.6-95$\,M$_\odot$. LVK inferred an intrinsic BH mass distribution for more massive BHs in BH-BH mergers. The distribution decreases steeply with the primary BH mass and it shows two peaks at $\sim $8\,M$_\odot$ and $\sim $35\,M$_\odot$. Understanding such result is a complex task.
The number of evolutionary channels may lead to the formation of BH-BH mergers (e.g. \citealt{Mandel2022}), and each channel has its own inherent uncertainties 
\citep[e.g., ][]{Belczynski2022}.
In many of the formation channels stars are the building blocks for BH-BH mergers. There is therefore the need for precise predictions from stellar evolution in regard to final state of massive stars at the time of core-collapse. Such predictions allow, among other things, to estimate stellar-origin BH masses. The BH mass is predicted to depend sensitively on metallicity~\citep{Belczynski2010a,Belczynski2010b}. Yet, a good starting point for such calculations is solar metallicity, because it represents a condition at which the effect of wind mass loss is highly noticeable, since many of the stellar wind models prescribe wind mass loss rates as a function of chemical abundances and luminosity, which both depend on metallicity levels. Additionally, solar metallicity is one of the most studied and better constrained cases. For instance, the most massive stellar BH found in the Milky Way is Cyg X-1, with $M_\text{BH}\simeq21.2\pm2.2$\,M$_\odot$ \citep{MillerJones2021}, being this value the lower threshold for theoretical estimations of $M_\text{BH,max}$.

Recently, two studies (that we are using as a comparison) offered new results in terms of final masses of massive stellar models. The final star mass may be taken as a proxy for BH mass (direct BH formation) or as an upper limit on BH mass (if there is a supernova explosion involved in the BH formation). First, ~\cite{Bavera2023} performed stellar evolution of massive stars (both as single stars and as parts of a binary system) in the Zero-Age Main Sequence (ZAMS) star mass range $M_{\rm ZAMS}=15-150$\,M$_\odot$ at $Z=0.014$ with the {\tt MESA} evolutionary code. They found that the BH mass from single star evolution increases from $M_{\rm BH}=3$\,M$_\odot$ at $M_{\rm ZAMS}=15$\,M$_\odot$ to a peak value at $M_{\rm BH}=35$\,M$_\odot$ at $M_{\rm ZAMS}=75$\,M$_\odot$, and then BH mass rapidly decreases with initial mass to $M_{\rm BH}=15$\,M$_\odot$ at $M_{\rm ZAMS}=100$\,M$_\odot$. They also found that for more massive stars ($M_{\rm ZAMS}>100$\,M$_\odot$) the BH mass does not change and remains more or less constant ($M_{\rm BH}=14$\,M$_\odot$). Second, ~\cite{Martinet2023} provided models performed with the {\sc Genec} (Geneva evolution code). We only pick the rotating models at $Z=0.014$ from this study for our comparison. These include $M_{\rm ZAMS}=180,\ 250,\ $300\,M$_\odot$. Final masses (end of C-burning) of these models are $M_{\rm final}=42,\,28$, and 37\,M$_\odot$. Additionally these models at final stage are WC/WO spectral subtype stars. For such final masses we may expect that they approximately correspond to BH masses. On one hand for such high final masses supernova models predict direct BH formation~\citep{Fryer2012}, on the other hand pulsational pair-instability supernovae are predicted to start removing mass for helium cores more massive than $\sim 35-40$~M$_\odot$~\citep{Woosley2017}. 

Both of the above calculations have employed ``Dutch''-like wind mass loss prescriptions, which have been considered for years the standard setup in 1D stellar evolution. The Dutch wind prescription orbits around the use of three different mass loss models: one for thin winds (\citealt{vink01}), one for thick winds where the switch from thin to thick winds is placed on the mass fraction of surface hydrogen  $X_{\rm surf}>$~0.4 (\citealt{NugisLamers2000}) and one for dust-driven winds at effective temperature $T_\text{eff}<$~10 kK (\citealt{deJager1988}). However, recent years offered various detailed studies of massive stars and their mass loss through stellar winds (e.g., \citealt{kk17,Bjorklund2021,alex23a}). We take advantage of these new results to update our calculations. Additionally, in ~\cite{Bavera2023} only non-rotating models were used, and in \cite{Martinet2023} a rather modest overshooting was employed ($\alpha_{\rm ov} \sim 0.2$). 

Here we propose an updated collection of wind-mass loss prescriptions that can be applied to all massive stars that are expected to form BHs. We also advocate for the use of rotating models to better represent the reality of stellar populations and the adoption of high overshooting values ($\alpha_{\rm ov}\sim0.5$) for massive stars as showed by ~\cite{Scott2021}. We perform most of our models with our updated version of {\tt MESA}, and we cross-check our {\tt MESA} results with a couple of calculations performed with our {\sc Genec} models. Both codes are made to run possibly the same physics (same winds, rotation, convection/overshooting, angular momentum and chemical transport). 

\begin{figure*}
\hspace*{-0.4cm}
\centering
\includegraphics[height=6 cm,width=0.9\textwidth]{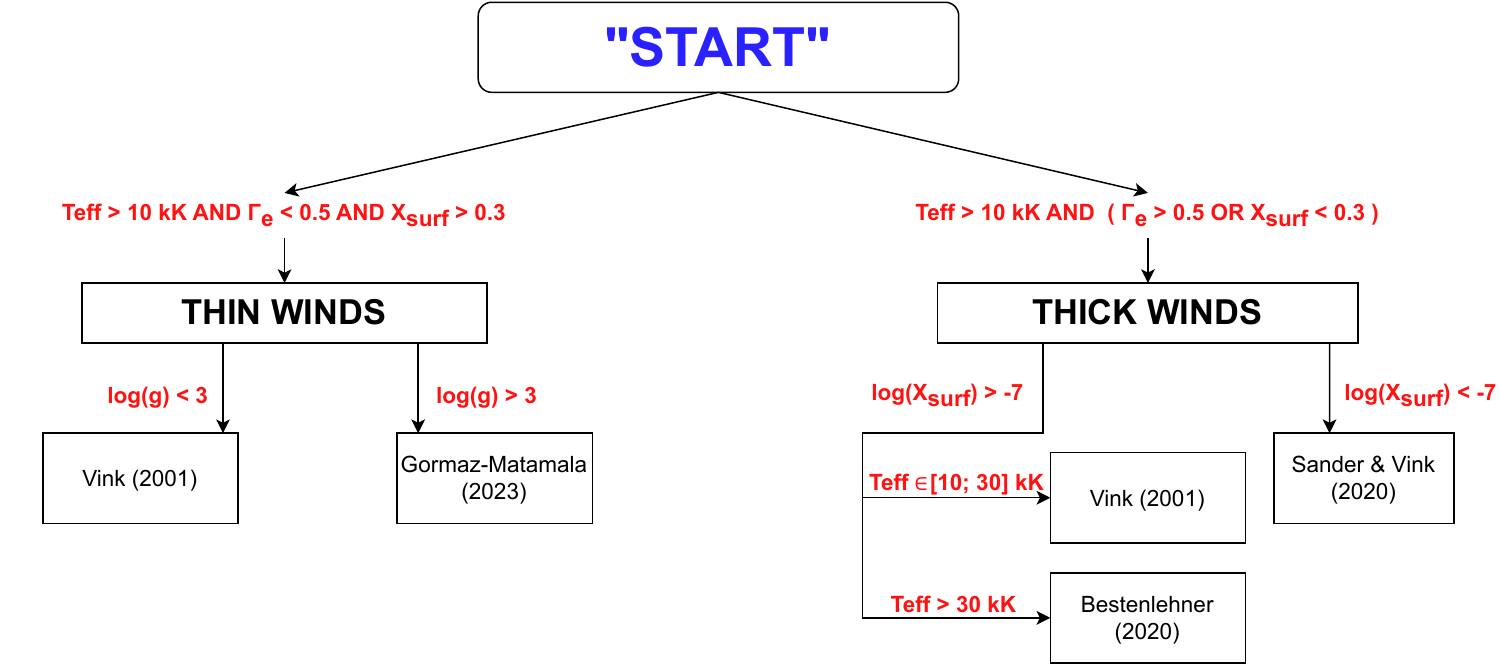}
\caption{
Adopted line-driven wind mass loss prescription for massive stars at solar metallicity (Z=0.014). For stars below 10 kK dust-driven winds will prevail and we will use \cite{deJager1988}.
See Sec.~\ref{sec.winds} for details. 
}
\label{fig.winds}
\end{figure*}

\section{Modeling} 
\label{sec.models}

In our study we use two different detailed evolutionary codes, the version 23.05.1 of Modules for Experiments in Stellar Astrophysics~\citep[\texttt{MESA}][]{Paxton2011,Paxton2013,Paxton2015,Paxton2018,Paxton2019,Jermyn2023}, and {\sc Genec} \citep{eggenberger08} for stars at solar metallicity $Z=0.014$~\citep{Asplund_2009} within an initial mass range between $10$ and 300~M$_\odot$. We unified the key input physics in both codes. We stop all our simulations at the depletion of carbon in the stellar core, since the timescale between the end of core C burning and the core-collapse is negligible in terms of star and core mass evolution (see e.g. \citealt{Woosley_2002}), and it does not therefore change the final estimates for BH masses.

\subsection{Mass loss prescription}
\label{sec.winds}

The theoretical calculation of wind mass loss rates ($\dot M$) depends on the physical properties of a star, and stellar winds change with the spectral type and evolutionary stage changes. In Figure \ref{fig.winds} we show our adopted wind mass loss prescription. Our choices are motivated by the recent work on various phases of wind mass loss for massive stars. Yet we use older prescriptions in parts of the parameter space where no updates are available or when the change from an older prescription to a more recent one would not be sufficiently motivated.

\begin{figure*}
\hspace*{-0.4cm}
\includegraphics[width=1.0\textwidth]{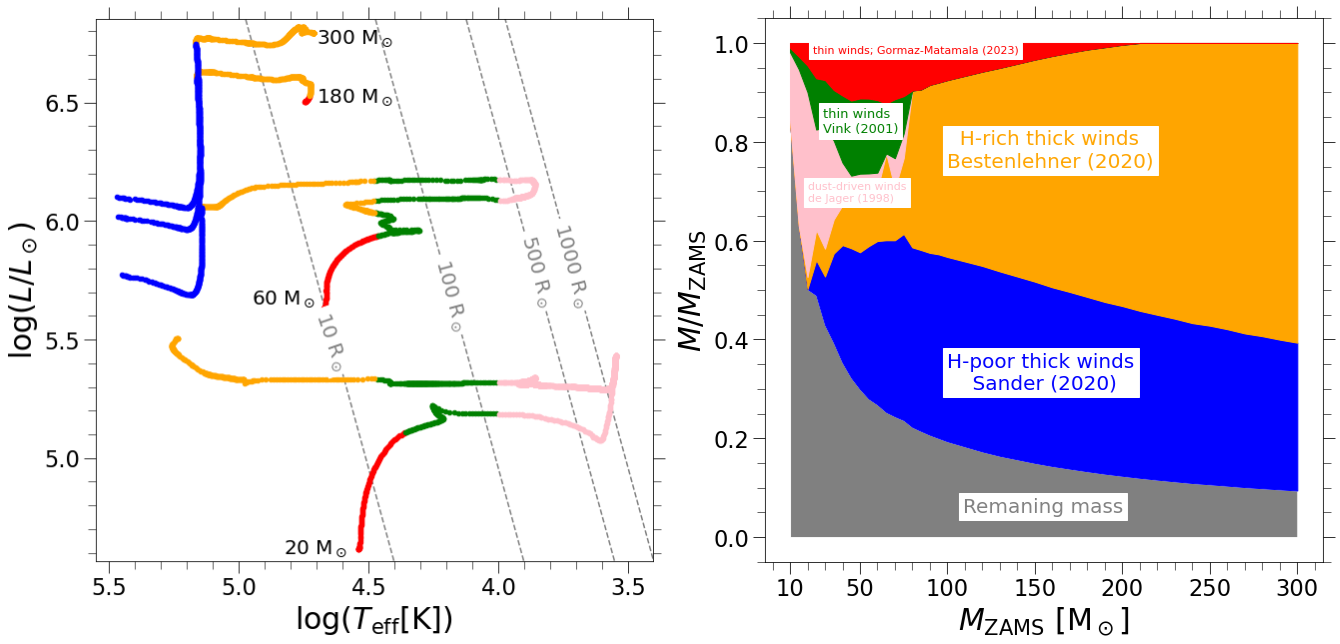}
\caption{
{\em Left:} H-R diagram for selected {\tt MESA} models. Colors correspond to various wind mass loss prescriptions used in our study. Note that stars with initial masses $M_{\rm ZAMS} \lesssim 60$~M$_\odot$ are subject to significant radial expansion, while more massive stars do not expand during evolution. {\em Right:} Fraction of ZAMS mass lost through given wind mass loss prescription (color coded as shown in the figure). Note that majority of mass loss for most massive stars 
($M_{\rm ZAMS} \gtrsim 60$~M$_\odot$) is expected by H-rich thick winds and by H-poor thick winds (for which we use recent updated formulae). 
}
\label{fig.hr}
\end{figure*}

\begin{figure*}
\centering
\includegraphics[width=0.95\textwidth]{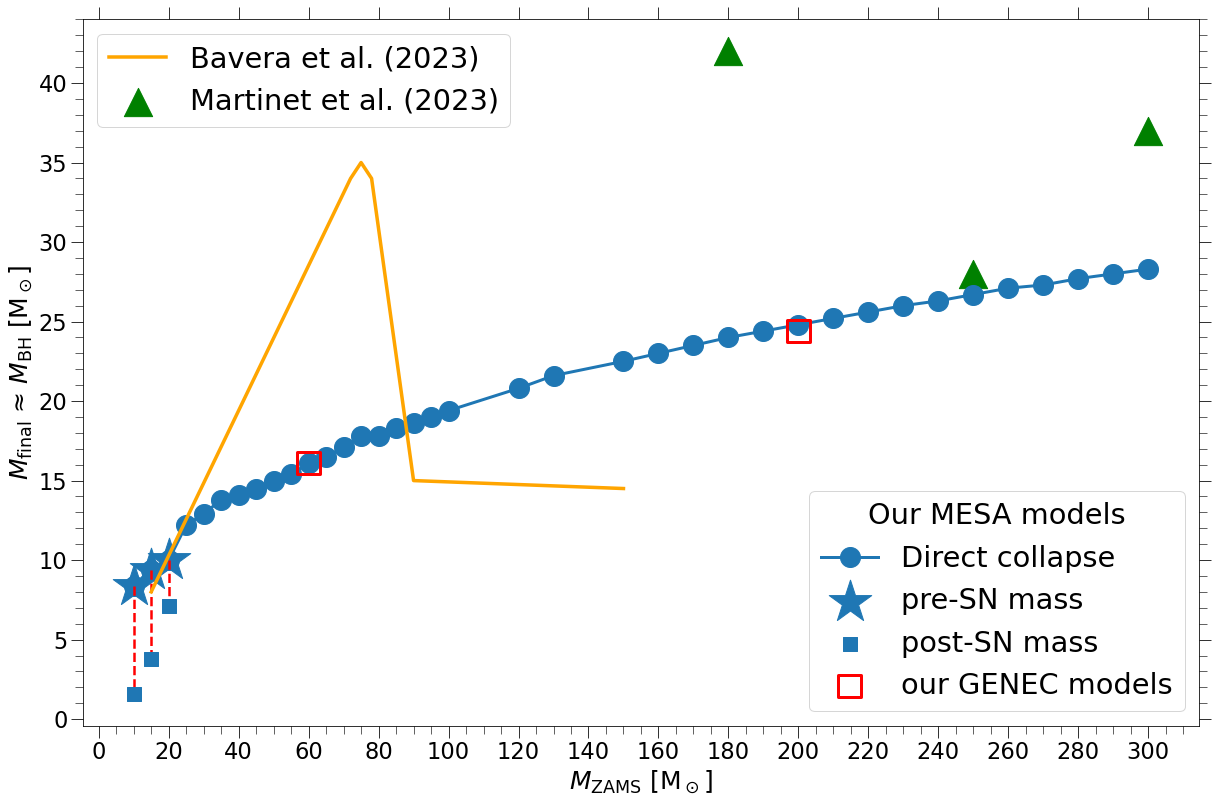}
\caption{
ZAMS mass --- final star mass (end of C-burning) relation at solar metallicity $Z=0.014$. Our updated {\tt MESA} models are shown with blue points (final mass of a star equals mass of a BH: no supernova) and blue stars (upper limit on mass of a BH, where blue squares show our estimates of NS/BH mass after supernova). Our two updated {\sc Genec} models are shown with red squares. Predictions from both codes match. We predict a monotonic growth of the BH mass as a function of its ZAMS mass. The predicted maximum BH mass in our models is $M_{\rm BH,max}=28.3$~M$_\odot$ and comes from a ZAMS mass $M_{\rm ZAMS}\sim 300$~M$_\odot$. As a comparison we also show the models from ~\cite{Bavera2023} for single star evolution (yellow line) and from ~\cite{Martinet2023} (green triangles): note that these models predict considerably different BH masses than our models. 
}
\label{fig.bhmass}
\end{figure*}

For line-driven winds, which are the dominant winds for the mass range we are evaluating in this work, we make the division between optically thin and optically thick winds.

Optically thin winds models imply that the stellar radius is well defined (such as OBA-type stars), whereas optically thick wind implies that there is not a clear separation between the photosphere 
and the expanded atmosphere, such as in Wolf-Rayet (WR) stars. For stellar evolution models, this distinction is made when the star is hot and $X_{\rm surf}$ is below certain value, normally $X_{\rm surf}<0.3$~or~$<0.4$ \citep{Yusof2013,kohler15}.
However, very massive stars such as WNh develop thick winds in early stages of their evolution \citep{tehrani19,martins22} because of their proximity to the Eddington limit due to their extreme luminosity \citep{vink11,bestenlehner14,grafener21}.
For this reason, for stars with large hydrogen mass fraction at surface we set the transition from thin to thick winds when the Eddington factor $\Gamma_\text{e}$ reaches some $\Gamma_\text{e,trans}$.
Studies analyzing the wind efficiency have found a smooth transition between thin and thick winds, without abrupt jumps in $\dot M$ \citep{vink12,Sabhahit2023}, but using enhanced values for the mass loss in the low $\Gamma_\text{e}$ regime.
For this work, we set $\Gamma_\text{e,trans}=0.5$, based on $\Gamma_\text{e,trans}\simeq0.47$ found by \citet{Bestenlehner2020}.

For optically thin winds we adopt the mass-loss rate from \citet{alex22b}, based on their self-consistent m-CAK prescription \citep{alex19,alex22a}.
	\begin{align}\label{mdotalex22bnew}
		\log\dot M_\text{GM23}=&-40.314 + 15.438\,w + 45.838\,x - 8.284\,w\,x \nonumber\\
		&+ 1.0564\,y -  w\,y / 2.36 - 1.1967\,x\,y\nonumber\\
		&+z\times\left[0.4+\frac{15.75}{M_*/{\rm M}_\odot}\right]\;,
	\end{align}
	where $\dot M_\text{GM23}$ is in M$_\odot$ yr$^{-1}$; and where $w$, $x$, $y,$ and $z$ 
        are defined as
	\begin{equation}
		w=\log \left(\frac{T_\text{eff}}{\text{kK}}\right)\;,\nonumber\\
		x=\frac{1}{\log g}\;,\\
		y=\frac{R_*}{R_\odot}\;,\\
		z=\log \left(\frac{Z_*}{Z_\odot}\right)\;.
	\end{equation}

The range of validity of m-CAK prescription is $\log g$~$\ge$~3.2 and $T_\text{eff}\ge$~30\,kK 
\citep{alex22a}. However, if we extend $\dot M_\text{GM23}$ for compact and cooler temperatures we 
can easily match with other mass-loss recipes such as \citet{kk17,kk18} at cooler stars, reason why we 
can revise the limit of validity as $\log g \ge 3.0$. Hence, $\dot M_\text{GM23}$ covers the most part 
of the Main Sequence (MS) and a fraction of post-MS expansion.
For the cases of $\log g<3.0$, we keep the classical formula of \citet[][hereafter Vink's formula]{vink00,vink01}, thus keeping the bi-stability jump in the mass loss around $25$ kK \citep{Vink1999} only for evolved stars. The existence of a jump in the regime of BSGs has been supported by recent theoretical \citet{Krticka2021,kk24} and empirical \citep{BerniniPeron2023} mass-loss studies.

For optically thick winds, i.e., when $X_\text{surf}<0.3$ or $\Gamma_\text{e}\ge0.5$, we use the 
$\dot M\propto\Gamma_\text{e}$ relationship calculated by \citet{bestenlehner20} for WNh stars.
	\begin{equation}\label{mdotbrands22}
		\log\dot M_\text{R136}=-5.19+2.69\log(\Gamma_\text{e})-3.19\log(1-\Gamma_\text{e})\;.
	\end{equation}

The constants stem from the calibration done over R136 by \citet{brands22}, which is located in the LMC ($Z=0.006$) and thus is valid for that metallicity only.
Therefore, to make Eq.~\ref{mdotbrands22} applicable for different metallicities, we include an extra term to readapt Bestenlehner's formula for $Z=0.014$.
	\begin{equation}\label{mdotbestenlehner20}
\log\dot M_\text{B20}=\log\dot M_\text{R136}+\log \left(\frac{Z_*}{Z_\text{LMC}}\right)\times\left[0.4+\frac{15.75}{M_*/{\rm M}_\odot}\right]\;,
	\end{equation}
{The extra metallicity dependent is the same as the term introduced in Eq.~\ref{mdotalex22bnew}, normalized for $Z_\text{LMC}$.
This formula is valid only for $T_\text{eff}\ge30$ kK, reason why for cooler stars we kept the validity of Vink's formula.

For H-poor regime ($X_\text{surf}\le10^{-7}$), we adopt the formula from \citet{sander20}, based on 
the hydrodynamically consistent wind calculations for classical WR stars \citet{sander20a}.
    \begin{equation}\label{sandervink20}
        \log\dot M=a\log[-\log(1-\Gamma_\text{e})]-\log 2\left(\frac{\Gamma_\text{e,d}}{\Gamma_\text{e}}\right)^{c_{d,b}}+\log\dot M_\text{off}\;,
    \end{equation}
where
$$a=2.932\;,$$
$$\Gamma_\text{e,b}=-0.324*\log \left(\frac{Z_*}{Z_\odot}\right)+0.244\;,$$
$$c_{d,b}=-0.44*\log \left(\frac{Z_*}{Z_\odot}\right)+9.15\;,$$
$$\log\dot M_\text{off}=0.23*\log \left(\frac{Z_*}{Z_\odot}\right)-2.61\;.$$

Hydrodynamically consistent solutions from \citet{sander20} have been recently upgraded by adding a temperature dependent term \citep{Sander2023}.
That term however, corresponds to the temperature calculated for a stellar radius which is not the same radius for WR stars adopted by \textsc{Genec} \citep{Meynet2005}.
Therefore, we adopt Eq.~\ref{sandervink20} without the temperature dependence, because its incorporation for stellar evolution models is beyond the scope of this work. Additionally, for the sake of consistency we also didn't adopt the temperature dependence in our {\tt MESA} setup. We nevertheless acknowledge that the addition of this temperature-dependent term represents an improvement for the modeling of H-poor WR stars, and future publications with the presented {\tt MESA} models will incorporate it in our wind-driven mass loss prescription.

\subsection{Evolutionary Codes}

For \texttt{MESA}, we adopt the Ledoux criterion for convective boundaries \citep{Ledoux_1947} and mixing length $l= 1.82$ \citep{Choi2016}. We use step-overshooting, with a value of $\alpha_{\rm ov}$ above every convective region that represents a modified version of the {\tt POSYDON} prescription described in \cite{Fragos2023}. For $M_{\rm ZAMS}<$~4~M$_\odot$ we use  $\alpha_{\rm ov}=0.16$, which was adopted in the {\tt MIST} project from calibrations of the Sun and open clusters \citep{Choi2016}. For 8~M$_\odot$~$<M_{\rm ZAMS}<$~20~M$_\odot$ we use $\alpha_{\rm ov}=0.415$ from the \cite{Brott2011} work.  Between 4~M$_\odot$ and 8~M$_\odot$ at ZAMS we interpolate $\alpha_{\rm ov}$ between 0.16 and 0.415. Finally, for $M_{\rm ZAMS}\ge$~20~M$_\odot$ we adopt $\alpha_{\rm ov}=0.5$ \citep{Scott2021}. Below each convective region we also adopt a value of undershooting $\alpha_{\rm under}$ that is $\alpha_{\rm under}=\alpha_{\rm ov}/5$. In order to reduce superadibaticity in regions near the Eddington limit we use the \textit{use\_superad\_reduction} method, which is stated to be a more constrained and calibrated prescription than \textit{MLT++} in {\tt MESA} \citep{Jermyn2023}. For each star we adopt an equatorial velocity over critical velocity at ZAMS equal to $\Omega/\Omega_{\rm crit}=0.4$\footnote{We show in Appendix~\ref{sec.appendix2} our parameter study where we analyzed the differences for BH masses arising from the implementation of different values of $\alpha_{\rm ov}$ and $\Omega/\Omega_{\rm crit}$ in our {\tt MESA} models.}. To model the rotation and mixing of material we use the calibrations from \cite{Heger_2000}, with the addition of Tayler-Spruit magnetic field-driven diffusion \citep{Tayler_1973,Spruit_2002,Heger_2005}.

For \textsc{Genec} we also employ the Ledoux criterion as \citet{georgy14} and \citet{sibony23} for convective boundaries. The treatment of rotation and its respective impact on mass loss is given by \citet{maeder00}, with  initial rotation of our models: $V_\text{rot}/V_\text{crit}=0.4$. The transport of angular momentum  inside the star follows the prescription of \citet{zahn92} complemented by \citet{maeder98}, whereas  here we use Taylor-Spruit dynamo following \citet{maeder04}. Same as for \texttt{MESA} we also use $\alpha_\text{ov}=0.5$, whereas the abundances correspond to the values given by \citet{Asplund_2009}, whereas opacities come from \citet{iglesias96}.

\section{Results}

The implementation of $\Gamma_{\rm e}$ as a complementary condition for the initiation of thick winds translates into increased mass loss rates for stars above 60~M$_\odot$ at ZAMS. As noticeable in Figure~\ref{fig.hr}, thick winds tend to be initiated comparatively earlier in the evolution of a star the higher its $M_{\rm ZAMS}$ is. The most massive stars almost spend no time at all in their thin winds phase, which leads them to promptly become WNh stars without any prior giant phase. This could potentially explain the lack of red supergiants (RSG) beyond the Humphrey-Davidson limit~\citep{Humphreys1994}. This comes in agreement with the conclusions of \cite{Mennekens2014}, who reached a similar result through the alteration of the mass loss rates during the Luminous Blue Variable (LBV) phase. We nevertheless highlight that the difference between our results and the ones we cite arises from the fact that our results attribute the WNh winds mass loss as the prohibitor of massive RSGs to being observed beyond the Humphrey-Davidson limit, while \cite{Mennekens2014} suggested it was due to the combined effect of winds mass loss and LBV outbursts.

The earlier transition from thin to thick winds at $\Gamma_\text{e}=0.5$ (instead of the previous $X_\text{surf}=0.3$ or $X_\text{surf}=0.4$) makes our new evolutionary tracks to find stars with optically thick envelopes and still a large fraction of hydrogen abundance, as the WNh stars observed in the Arches cluster \citep{Martins2008,Martins2023}. We also show that the proposed mass-loss prescription leads every star at $M_{\rm ZAMS} \gtrsim 20$~M$_\odot$ (roughly the minimum initial mass of BH progenitors at $Z=0.014$ from single star evolution) to evolve into a Wolf-Rayet star during the latest stages of its evolution. Below that initial mass, the mass loss rates and the internal mixing are not enough for stars to enter the thick winds regime. We highlight that this scenario may change depending on the adopted prescription for dust-driven winds (e.g. \citealt{vanLoon2005,GonzalezTora2023,GonzalezTora2023b}). For instance \cite{Beasor2021,Beasor2023} show considerably lower mass loss rates than \cite{deJager1988} and more specifically \cite{Beasor2023} claims that winds under this regime ``are inefficient at removing the H envelope".

We report the final masses of stars as a function of their initial mass at solar metallicity calculated with {\tt MESA} and {\sc Genec} in Figure~\ref{fig.bhmass}. We have unified the input key physics in both codes in terms of rotation, convection, overshooting, and wind mass loss rates. We did not calibrate the codes any further for the results to match. Yet, for the two test models (at $M_{\rm ZAMS}=60$~M$_\odot$ and $200$~M$_\odot$) both codes produce almost identical final star masses that show robustness of our predictions.  We also show as a comparison the estimates from \cite{Martinet2023} and an approximation of the BH mass distribution from \cite{Bavera2023}. Most of the stars in our models, according to \cite{Fryer2012}, collapse directly into BHs without any SN event.

In contrast with other models like \cite{Bavera2023} we do not report a quasi-flat distribution of BH masses for the most massive stars in our models, nor a BH mass peak in the low mass range. On the contrary, the BH mass increases monotonically as a function of ZAMS mass to $M_{\rm BH}=27.7$~M$_\odot$ for our most massive model at $M_{\rm ZAMS}=300$~M$_\odot$. On the other hand, our models predict lower final star masses than the ones from \cite{Martinet2023} for the highest ZAMS masses. This is due to the fact that in canonical {\sc Genec} simulations the switch from thin to thick winds is based only on H surface abundance, plus the upgrade of the mass loss prescription for the late evolutionary stages \citep{Higgins2021}. 
This makes even the most massive stars to spend significant part of the lifetime in thin (low) wind mass loss regime. 
In our models very massive stars enter the thick (high) wind mass loss regime very early in the evolution (note that we also use the Eddington factor in addition to surface abundance for the switch). As pointed out in Section~\ref{sec.winds} the upgraded switch using both H surface abundance and the Eddington factor seems in better agreement with observations.

\section{Conclusion}

Our study shows that in solar metallicity environments, assuming the current defaults for dust-driven mass loss, most of the black hole progenitors in their latest stages of their evolution naturally become Wolf-Rayet stars even without invoking tidally-induced rotational mixing or a mass transfer event with a stellar companion. This is due to the combined effect of internal mixing and strong stellar winds at high metallicity levels.
Our results substantially differ from recent results in the context of black hole masses at solar metallicity. We find a different initial--final star mass relation for our updated stellar models. 
Firstly, our models predict a maximum black hole mass of $\sim 28$~M$_\odot$ that is smaller by at least $\sim 7$~M$_\odot$ if compared to~\cite{Bavera2023} and ~\cite{Martinet2023}. Secondly, our most massive black holes are formed from significantly different initial star masses ($M_{\rm ZAMS}>250$~M$_\odot$) than the ones found by \cite{Bavera2023} from single star evolution~($M_{\rm ZAMS} \sim 75$~M$_\odot$). These differences are understood in terms of the updated wind mass loss prescriptions and of the internal mixing prescription (see Appendix \ref{sec.appendix2} for more details on the role of overshooting and rotational velocity). Note that we have not only adopted different wind mass loss prescriptions but we have also proposed transition criteria among prescriptions that are more complex than traditionally adopted in the modeling of massive stars. Following these results we claim that the formation of a black hole with a mass above 27~M$_\odot$ can likely only come from a) interaction with other objects, be it accretion from interstellar medium, mass transfer event with a star or merger with other stellar/compact objects b) the Big Bang (i.e. primordial black holes) c) the isolated evolution of a star with $M_{\rm ZAMS}\gtrsim250$~M$_\odot$. 

The shape of the initial-final star mass relation at solar metallicity is an important calibration point in the predictions of the overall black hole mass distribution in Universe. LIGO/Virgo/Kagra observatories are sensitive to detecting black holes from a wide range of metallicities (i.e., already reaching galaxies to redshift of $z\sim 1$). Such studies as presented here need to be extended to a full range of metallicities before any reliable initial mass conclusions can be made about the black hole mass distribution observed in gravitational-waves. For lower metallicities wind mass loss rates are smaller and final star masses are larger than predicted for $Z=0.014$. This will introduce another key factor/uncertainty in black hole mass calculation: pair-instability pulsations mass loss and stellar disruptions \cite[e.g.,][]{Farmer2020}.

We are not claiming that our calculations represent a definitive answer to the question of black hole mass distribution at solar metallicity. Yet, we argue that given the employed recent advancements in massive stellar evolution, this is an incremental step forward in understanding stellar-origin black holes.  

\vspace*{-0.2cm} \acknowledgements
Authors acknowledge support from the Polish National Science Center grant Maestro (2018/30/A/ST9/00050). We would also like to thank number of people for their helpful and/or critical comments on our modelling: Raphael Hirschi, Sylvia Ekström, Georges Meynet, Chris Fryer, Jakub Klencki, Simone Bavera, Tassos Fragos, Joachim Bestenlehner and Gemma Gonzalez-Tora. The authors also thank the {\tt MESA} community at large for their helpful feedback on the models creation. Computations for this article have been performed using the computer cluster at CAMK PAN.
We dedicate this paper to Krzysztof Belczyński, who contributed to this research before his untimely passing on the 13$^{\rm th}$ of January 2024.

\bibliography{ms}


\begin{appendix}

\section{Appendix A: Comparison of old and new wind mass loss prescriptions}
\label{sec.appendix}

The most remarkable difference between old and new wind prescriptions is the value of the mass-loss 
rate. These lower values of $\dot M$ show a better agreement with the current understanding of stellar winds, 
which are known to be clumped \citep{Bouret2005,Surlan2012,Surlan2013}.
More detailed theoretical studies on the wind structure of O-type stars, self-consistently coupling radiative acceleration and hydrodynamics in the co-moving frame and non-LTE conditions \citep{Sander2017,alex21}, have confirmed the same diagnostics.

For thin winds, we observe in Figure~\ref{fig.thinthick} that the self-consistent $\dot M$ from the m-CAK prescription \citet{alex23a} agrees very well with the formula from \citet{kk17,kk18} especially for MS stars as outlined by \citet{alex22a}. Wind solution from \citet{Bjorklund2021} also is lower than Vink's formula, although their $\dot M$ is extremely low at low luminosities, together with not evidencing the increase of $\dot M$ around $\sim25$ kK due to the bi-stability jump.
Whereas for thick winds, we observe in Figure~\ref{fig.thinthick} that both \citet{bestenlehner20} and \citet{sander20} provide mass-loss rates below the former \citet{NugisLamers2000}. Mass loss for H-rich stars is lower, because it describes the winds of stars in an intermediate evolutionary stage between O-type and WR stars.

However, notice that both panels of Figure~\ref{fig.thinthick} only show the relationship $\dot M\sim L_*$, and straightforwad differences in mass-loss rates at the same luminosity may be misleading. For example lowering mass-loss rates makes stars reach higher luminosities during their evolution~\citep{alex22b,alex23a,Bjorklund2023}, and therefore the mass-loss rates at subsequent evolutionary stages may be larger rather than smaller after adopting updated ``lower'' wind mass loss prescriptions.
As another example, $\dot M$ for H-rich WR stars \citep{Bestenlehner2020} is lower than $\dot M$ for H-poor stars \citep{sander20} at same luminosity, because they describe stars with different $L/M$ ratio (and thus different mass regimes).

\begin{figure*}
\hspace*{-0.4cm}
\includegraphics[width=1.0\textwidth]{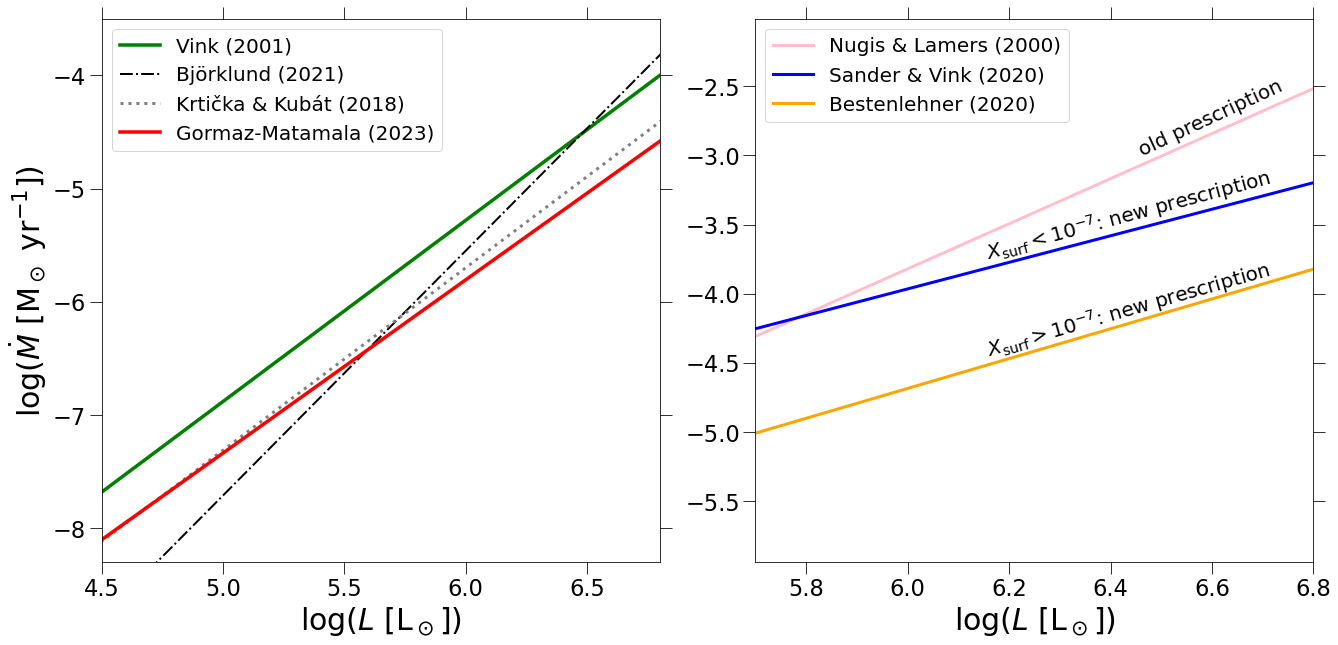}
\caption{
Comparison of old and new wind prescriptions for thin (left panel) and thick
(right panel) winds.
}
\label{fig.thinthick}
\end{figure*}

\section{Appendix B: Comparison between different overshooting values and initial rotational velocities}
\label{sec.appendix2}

We show in Figure~\ref{fig.merge_Rot_Oversh} our parameter study for different initial rotation velocities (while keeping $\alpha_{\rm ov}$\,=\,0.5) and convective overshooting parameter (with the initial $\Omega/\Omega_{\rm crit}$\,=\,0.4). Within the considered parameter space a more efficient mixing brings stars to enter their thick winds and/or dust-driven winds phases earlier in their evolution, which nullifies any noticeable difference in terms of helium core mass at the end of the MS. This explanation gains also a stronger physical motivation for very massive stars $M_{\rm ZAMS}\gtrsim$100\,M$_\odot$, which are almost fully convective (i.e. they do not need overshooting nor rotation to mix elements throughout their whole volume) and enter the thick winds regime almost as early as ZAMS. As a reference, we also show some extreme cases on the r.h.s. plot of Figure~\ref{fig.merge_Rot_Oversh} with very inefficient mixing within stellar interiors (no rotation and $\alpha_{\rm ov}\leq$0.2), if compared with our reference model described in Section \ref{sec.models}. We report that for non-rotating stars there is no smooth transition as a function of $\alpha_{\rm ov}$ from the $M_{\rm final}$ behavior with a local maximum at $M_{\rm ZAMS}\sim$60\,M$_\odot$ shown for $\alpha_{\rm ov}\leq$0.15 to the monotonic behavior that is noticeable in all other models. We explain that this drastic transition is due to the fact that for these models with relatively low mixing efficiencies stars do not reach the WR phase. To support this statement we show Table \ref{tab:table}, where we report the cumulative mass lost from the adopted wind mass loss prescriptions for a 60\,M$_\odot$ star (roughly the $M_{\rm ZAMS}$ where we see the localized $M_{\rm final}$ maximum for the lower mixing efficiency cases) for different initial conditions of $\alpha_{\rm ov}$ and $\Omega/\Omega_{\rm crit}$.


\begin{figure*}
\centering
\includegraphics[width=\textwidth]{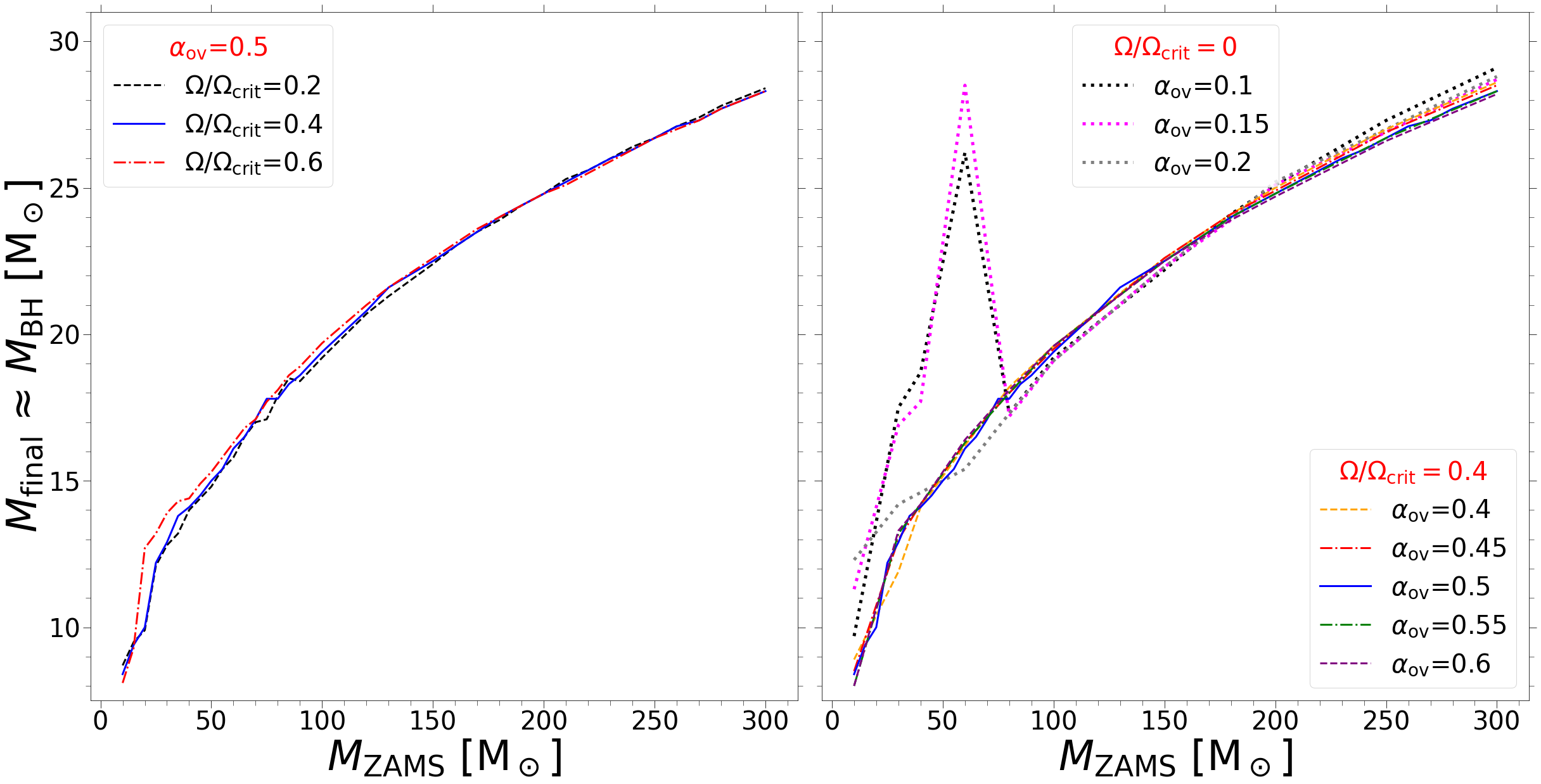}
\caption{ZAMS mass -- final star mass (end of core C-burning) relation at Z$_{\rm \odot}$. On the l.h.s plot we compare the results of our simulations between different initial rotation velocities at ZAMS. On the r.h.s. plot, instead, we compare the results from the simulations made with different values for the convective overshooting parameter $\alpha_{\rm ov}$. The dotted lines represent non-rotating stars with $\alpha_{\rm ov}\leq$0.2.
}
\label{fig.merge_Rot_Oversh}
\end{figure*}

\begin{deluxetable}{c|cccccc}
\tablecaption{Mass lost by a 60 M$_\odot$ star at Z$_\odot$ from each and all the adopted wind-mass loss prescriptions at different initial conditions. \label{tab:table}}
\tablehead{
\colhead{Model} & \colhead{$\Delta M_{\rm GM23}$} & \colhead{$\Delta M_{\rm V01}$} & \colhead{$\Delta M_{\rm dJ88}$} & \colhead{$\Delta M_{\rm B20}$} & \colhead{$\Delta M_{\rm SV20}$} & \colhead{$\Delta M_{\rm TOT}$} \\
\colhead{} & \colhead{[M$_\odot$]} & \colhead{[M$_\odot$]} & \colhead{[M$_\odot$]} & \colhead{[M$_\odot$]} & \colhead{[M$_\odot$]} & \colhead{[M$_\odot$]}
}
\startdata
$\alpha_{\rm ov}=0.1 \hspace{0.1 cm};\hspace{0.1 cm} \Omega/\Omega_{\rm crit}=0$ & 5.2 & 13.5 & 15.0 & 0 & 0 & 33.7\\
$\alpha_{\rm ov}=0.15 \hspace{0.1 cm};\hspace{0.1 cm} \Omega/\Omega_{\rm crit}=0$ & 5.0 & 10.5 & 15.8 & 0 & 0 & 36.3\\
$\alpha_{\rm ov}=0.2 \hspace{0.1 cm};\hspace{0.1 cm} \Omega/\Omega_{\rm crit}=0$ & 5.0 & 5.4 & 17.1 & 4.4 & 12.7 & 44.6\\
$\alpha_{\rm ov}=0.5 \hspace{0.1 cm};\hspace{0.1 cm} \Omega/\Omega_{\rm crit}=0.4$ & 7.2 & 8.9 & 2.1 & 5.8 & 19.8 & 43.8
\enddata
\tablecomments{GM23: \cite{alex23a}; V01: \cite{vink01}; dJ88: \cite{deJager1988}; B20: \cite{bestenlehner20}; SV20: \cite{sander20}; $\Delta M_{\rm TOT}$: total wind-driven mass lost }
\end{deluxetable}

\end{appendix}


\end{document}